\documentclass[aps,showpacs,prd,twocolumn,superscriptaddress,nofootinbib,preprintnumbers]{revtex4-1}

\usepackage{amsmath}
\usepackage{amsfonts}
\usepackage{amssymb}
\usepackage{graphicx}
\usepackage{epstopdf}
\usepackage{longtable}

\newcommand{\tr}{\mathrm{tr}\,}

\newlength{\pbln}
\newcommand{\pb}[1]{\parbox{\pbln}{#1}}

\begin{document}
\preprint{WUB/10-21}

\title{A numerical investigation of orientifold planar equivalence for quenched mesons}


\author{Biagio Lucini}
\email[]{b.lucini@swansea.ac.uk}
\affiliation{School of Physical Sciences, Swansea University
Singleton Park, Swansea SA2 8PP, UK}
\author{Gregory Moraitis}
\email[]{pygm@swansea.ac.uk}
\affiliation{School of Physical Sciences, Swansea University
Singleton Park, Swansea SA2 8PP, UK}
\author{Agostino Patella}
\email[]{a.patella@swansea.ac.uk}
\affiliation{School of Physical Sciences, Swansea University
Singleton Park, Swansea SA2 8PP, UK}
\author{Antonio Rago}
\email[]{rago@physik.uni-wuppertal.de}
\affiliation{Department of Physics, Bergische Universit\"at Wuppertal, Gaussstr. 20, D-42119 Wuppertal, Germany}


\date{\today}

\begin{abstract}
We measure on the lattice the quenched pseudoscalar and vector meson masses at a fixed value of the lattice spacing for SU($N$) gauge theory with fermions in the adjoint, in the symmetric and in the antisymmetric representation of the gauge group. Simulations are performed for $N=3,4,6$ in all those representations, with the addition of $N=2$ for the adjoint representation. We illustrate a strategy for separating the even from the odd-power contributions in $1/N$ in the masses. Using this technique, we extrapolate the vector mass to the large-$N$ limit in the chiral region and show that at $N = \infty$ this mass is the same within errors in all the three representations, as predicted by orientifold planar equivalence. Possible implications of our investigation for studying orientifold planar equivalence in the dynamical case are discussed. 
\end{abstract}

\maketitle

\section{Introduction}
\label{sec:intro}

Orientifold planar equivalence~\cite{Armoni:2003gp,Armoni:2003fb,Armoni:2003yv,Armoni:2004ub,Patella:2005vx} is the equivalence in the large-$N$ limit and in a common sector of the following theories: \textbf{sQCD}/\textbf{asQCD}, i.e. the SU($N$) gauge theory with $N_f$ Dirac fermions in the symmetric/antisymmetric two-index representations of the gauge group, and \textbf{adjQCD}, i.e. the SU($N$) gauge theory with $N_f$ Majorana fermions in the adjoint representation of the gauge group. The common sector is defined as the set of all the states and single-trace observables that are invariant under charge conjugation (C) symmetry. It was shown in~\cite{Unsal:2006pj,Unsal:2008ch} that orientifold planar equivalence holds if and only if the C-symmetry is not spontaneously broken in the three theories, which ensures that the vacuum is in the common sector.

On a general ground, orientifold planar equivalence is only one in a rich network of equivalences arising in the large-$N$ limit of gauge theories, which includes gauge group independence, orbifold equivalences and volume independence~\cite{Kovtun:2003hr,Kovtun:2004bz,Kovtun:2007py,Unsal:2008ch}. The implications of these equivalences are nowadays only partially understood and their investigation represents a fascinating challenge from an analytical point of view.

Orientifold planar equivalence is also an important tool for investigating real QCD. In fact it was observed~\cite{Corrigan:1979xf} that asQCD is a fully legitimate multicolor generalization of QCD, since for the real-world case $N=3$ the antisymmetric two-index representation is equivalent to the antifundamental one. A priori there is no reason for the standard 't~Hooft~\cite{Hooft:1973jz} large-$N$ generalization (i.e. all fermions in the fundamental representation) to be preferred to asQCD. A simple computation of the one-loop beta function shows that the standard 't~Hooft large-$N$ limit underestimates the role of the fermions (fermionic loops are suppressed), while the orientifold large-$N$ limit (obtained using asQCD) overestimates the role of the fermions. Thanks to orientifold planar equivalence, the orientifold large-$N$ limit of QCD is equivalent to large-$N$ adjQCD (which is super Yang-Mills in the case of one massless flavor).

Once established that adjQCD can be viewed as an approximation of QCD in the sense of the orientifold large-$N$ limit, it is natural to ask how good this approximation is, or in other words how large the $1/N$ corrections are. The $1/N$ corrections are generally beyond the reach of analytical tools, but they can be computed by means of numerical simulations. Extensive numerical studies of the $1/N^2$ corrections of the standard 't~Hooft large-$N$ limit exist in the literature (see e.g.~\cite{Teper:2009uf,Narayanan:2009xh} for recent reviews). Numerical simulations of SU($N$) gauge theories with dynamical fermions in the two-index representations are very time-consuming. The idea behind this work is to approach the fully dynamical simulations of these theories in steps of increasing difficulty, starting from the simpler case of the quenched theories. In the process, we can gather experience about fermions in the two-index representations and we shall develop the needed techniques to handle the $1/N$ corrections (as opposed to the $1/N^2$ corrections) which naturally appear in s/asQCD. In this spirit a first paper~\cite{Armoni:2008nq} appeared in which the fermionic condensate was computed. From that study, we learned that in adjQCD, for a wide range of bare masses, the fermionic condensate is well-described down to $N=2$ (within a few percent) just by including the first subleading correction (${\cal O}(1/N^2)$). In s/asQCD three subleading corrections (up to ${\cal O}(1/N^3)$) must be included in order to describe the data down to $N=3$, while $N=2$ is out of reach. Numerically, asQCD for low $N$ is the representation that is farthest from the common large-$N$ limit. In order to accurately determine the finite-$N$ corrections up to ${\cal O}(1/N^3)$ in s/asQCD, numerical data in the two representations were combined to separate the corrections corresponding to even and odd powers of $1/N$. This technique proved to be quite effective at determining the large-$N$ behavior from numerical studies ranging up to $N=6$.

In this paper we present a similar study for quenched isovector mesonic observables. We will present results for the pseudoscalar ($PS$) and vector ($V$) meson masses\footnote{For a recent analytical computation of those masses and a discussion of orientifold planar equivalence at strong coupling, see~\cite{Moraitis:2009xt}.}. Our conclusions will be qualitatively similar to the case of the fermionic condensate.

A comment is in order at this point. Although originally proved for theories with dynamical fermions, orientifold planar equivalence is valid also for quenched theories. A convenient way to define the quenched theories is to add ghost fields (bosons with a spinor index) that kill the fermionic determinant once they are integrated out. For instance on the lattice:
\begin{gather}
S_{Q} = S_{YM} + \sum_x \left\{ \sum_{f=1}^{N_f} \bar{\psi}_f D_m \psi_f(x) + \sum_{f=1}^{N_f} c_f^\dag D_m c_f(x) \right\} \ .
\end{gather}
The proof of orientifold planar equivalence is a simple extension of the one presented in~\cite{Armoni:2004ub,Patella:2005vx}, once the extra fields are included. It is worth noticing that in the dynamical case it is necessary to prove the equivalence of the vacua in the three theories in the large-$N$ limit, while in the quenched case the vacua are identical by construction for each value of $N$, since the vacuum of the system is always the pure Yang-Mills vacuum. In this sense orientifold planar equivalence is a less rich result in the quenched case. We should emphasize that for theories with two-index fermions the large-$N$ limit is not the quenched theory. Hence, our investigation must be regarded as a preliminary step towards the dynamical case.

This paper is organized as follows. In Sect.~\ref{sect:simulations} we present the details of our numerical investigation. Sect.~\ref{sect:corrections} shows how the contributions of the odd and even power terms in $1/N$ can be separated for the amplitudes and masses of correlation functions. Our numerical results are discussed in Sect.~\ref{sect:numerical}. Finally, our findings are summarized and discussed in Sect.~\ref{sect:conclusions}.

\section{Details of simulations}
\label{sect:simulations}
In the quenched theory the gauge field configurations are distributed as in pure Yang-Mills. We used $32 \times 16^3$ lattices, we studied the system for $N=3,4,6$ in all representations (with the addition of $N=2$ for the adjoint) and we performed simulations at the $\beta$ values corresponding to a deconfinement temperature $(5a)^{-1}$ (where $a$ is the lattice spacing)~\cite{Lucini:2002ku,Lucini:2003zr}. Lattice configurations were generated with the Wilson action and the Cabibbo-Marinari algorithm, as described e.g. in~\cite{DelDebbio:2007wk}, and correlators were measured using Wilson fermions. We used the HiRep code which has been developed for simulating generic number of colors and generic fermionic representation. Both the HiRep code and the Wilson-Dirac operator in a generic representation of the gauge group are described in~\cite{DelDebbio:2008zf}. 

The basic observables we measured are the isovector mesonic correlators. For s/asQCD, $u$ and $d$ are two flavours of Dirac fermions, we define
\begin{flalign}
& C_{S/AS}(t)
= \frac{1}{L^3} \sum_{\mathbf{x}} \langle 0 | (\bar{u} \Gamma d)^\dag(t,\mathbf{x}) (\bar{u} \Gamma d)(0,\mathbf{0}) | 0 \rangle = \nonumber \\
&= \frac{1}{L^3} \sum_{\mathbf{x}} \langle \tr [ D^{-1}_{S/AS}(0,\mathbf{0};t,\mathbf{x}) \Gamma^{\dag} D^{-1}_{S/AS}(t,\mathbf{x};0,\mathbf{0}) \Gamma ] \rangle \ ,
\end{flalign}
where $\Gamma=\gamma_5$ for the pseudoscalar ($PS$) channel and $\Gamma=\gamma_1$ for the vector ($V$) channel. For adjQCD, $u$ and $d$ must be considered as two flavors of Majorana fermions, and this gives rise to an extra $1/2$ factor:
\begin{flalign}
& C_{Adj}(t)
= \frac{1}{L^3} \sum_{\mathbf{x}} \langle 0 | (\bar{u} \Gamma d)^\dag(t,\mathbf{x}) (\bar{u} \Gamma d)(0,\mathbf{0}) | 0 \rangle = \nonumber \\
& = \frac{1}{2L^3} \sum_{\mathbf{x}} \langle \tr [ D^{-1}_{Adj}(0,\mathbf{0};t,\mathbf{x}) \Gamma^{\dag} D^{-1}_{Adj}(t,\mathbf{x};0,\mathbf{0}) \Gamma ] \rangle \ .
\end{flalign}

For extracting the masses from correlators we use the method described in~\cite{Fleming:2009wb} and summarized in the appendix of~\cite{DelDebbio:2010hu}, taking into account also the contribution of the first excited state. We refer to the $PS$ and $V$ isovector meson masses respectively as $m_{PS}$ and $m_V$. The values of the bare mass $m_0$ were chosen in the region in which the chiral behavior
\begin{gather}
\label{eq:chiral}
a m_V = a m_V^\chi + B (a m_{PS})^2 
\end{gather}
is visible. The location and extent of the chiral region depends on the value of $N$ and on the fermion representation.

\section{Separating even/odd contributions}
\label{sect:corrections}

From diagrammatic arguments, in pure Yang-Mills the Taylor expansion in $1/N$ of observables like normalized products of Wilson loops contains only even powers of $1/N$. We want to show that the same happens for isovector mesonic correlators in quenched adjQCD, while in quenched s/asQCD odd powers of $1/N$ also appear.

In fact using the hopping expansion, those correlators can be written as a sum of Wilson loops in the fermionic representation $R$:
\begin{gather}
C_R(t) = n_f^R \sum_{\mathbf{x}} \sum_{\omega} c_\omega \langle \tr R[W(\omega)] \rangle_{YM} \ ,
\end{gather}
where $\omega$ is a generic closed path going through the origin and the point $(\mathbf{x},t)$, $c_\omega$ is a coefficient independent of $N$ and the representation $R$, and $W(\omega)$ is the Wilson loop along the path $\omega$. $n_f^R$ is a coefficient that depends on the representation; in particular, we have $n_f = 1$ for the symmetric and antisymmetric representations and $n_f=1/2$ for the adjoint representation.

In addition to the adjoint, symmetric and antisymmetric representations, it is useful to consider also the reducible two-index representation ($F^2$). We will refer to the SU($N$) gauge theory with fermions in the reducible two-index representation as \textbf{f2QCD}. For these representations we can use the algebraic relationships:
\begin{equation}
\begin{split}
&\tr Adj[W] = |\tr W|^2 - 1 \ , \\
&\tr AS[W] = \frac{(\tr W)^2 - \tr (W^2)}{2} \ , \\
&\tr S[W] = \frac{(\tr W)^2 + \tr (W^2)}{2} \ , \\
&\tr F^2[W] = \tr S[W] + \tr AS[W] = (\tr W)^2 \ .
\end{split}
\end{equation}

Consider first the normalized correlator for quenched f2QCD:
\begin{flalign} \label{eq:f2corr}
\frac{C_{F^2}(t)}{N^2} = & \frac{C_{S}(t)+C_{AS}(t)}{N^2} = \nonumber \\
= & \sum_{\mathbf{x}} \sum_{\omega} c_\omega \frac{\langle [\tr W(\omega)]^2 \rangle_{YM}}{N^2} \ .
\end{flalign}
Since $\langle [\tr W(\omega)]^2 \rangle_{YM}/N^2$ contains only even powers in $1/N$, the same conclusion can be drawn for $C_{F^2}(t)/N^2$. One can be tempted to conclude that also the mesonic masses in quenched f2QCD must contain only even powers in $1/N$. We will show that this is not correct.

Consider separately the normalized correlators for quenched s/asQCD:
\begin{flalign}
\frac{C_{S/AS}(t)}{N^2} = & \frac{1}{2} \sum_{\mathbf{x}} \sum_{\omega} c_\omega \frac{\langle [\tr W(\omega)]^2 \rangle_{YM}}{N^2} + \nonumber \\
& \pm \frac{1}{2N} \sum_{\mathbf{x}} \sum_{\omega} c_\omega \frac{\langle \tr W(\omega\omega) \rangle_{YM}}{N}\ .
\label{eq:S/AScorr}
\end{flalign}
The second term clearly contribute with odd powers of $1/N$. A rapid inspection reveals that summing the correlators in quenched s/asQCD kills the odd powers in $1/N$. Expanding the correlators in eigenstates of the Hamiltonian gives
\begin{gather}
\frac{C_{S/AS}(t)}{N^2} = \sum_n a_{n,S/AS} \left( \frac{1}{N} \right) \exp \left\{ - t m_{n,S/AS} \left( \frac{1}{N} \right) \right\} \ .
\end{gather}
The masses and the amplitudes in quenched s/asQCD must contain both even and odd powers of $1/N$, which can be formally separated:
\begin{equation}
\begin{split}
&a_{n,S/AS} \left( \frac{1}{N} \right) = A_{n,S/AS} \left( \frac{1}{N^2} \right) + \frac{1}{N} \alpha_{n,S/AS} \left( \frac{1}{N^2} \right) \ , \\
&m_{n,S/AS} \left( \frac{1}{N} \right) = M_{n,S/AS} \left( \frac{1}{N^2} \right) + \frac{1}{N} \mu_{n,S/AS} \left( \frac{1}{N^2} \right) \ .
\end{split}
\end{equation}
We indicate with the same $n$ states in the $S$ and $AS$ channels which are related by orientifold planar equivalence:
\begin{equation}
\begin{split}
&a_{n,S}(0) = a_{n,AS}(0) = A_{n,S}(0) = A_{n,AS}(0) \ ,\\
&m_{n,S}(0) = m_{n,AS}(0) = M_{n,S}(0) = M_{n,AS}(0) \ .
\end{split}
\end{equation}

Once expanded in eigenstates of the Hamiltonian, the correlator for quenched f2QCD is:
\begin{flalign}
\frac{C_{F^2}(t)}{N^2} = & \sum_n a_{n,S} \left( \frac{1}{N} \right) \exp \left\{ - t m_{n,S} \left( \frac{1}{N} \right) \right\} + \nonumber \\
& + \sum_n a_{n,AS} \left( \frac{1}{N} \right) \exp \left\{ - t m_{n,AS} \left( \frac{1}{N} \right) \right\} \ .  \label{eq:f2_masses_corr}
\end{flalign}
Since the spectrum of quenched f2QCD is given by the union of the spectra of $S$ and $AS$ channels, masses and amplitudes in quenched f2QCD contain also odd powers of $1/N$. According to~\eqref{eq:f2corr}, the $F^2$ correlator cannot contain odd-power contribution in $1/N$, therefore these contributions in each term of~\eqref{eq:f2_masses_corr} must cancel each other. In formulae this means that the antisymmetrized combination in $1/N$ of the $F^2$ correlator must vanish:
\begin{flalign} \label{eq:oddpowers}
\frac{1}{N^2} & \left[ C_{F^2} \left( t, \frac{1}{N} \right) - C_{F^2} \left( t, -\frac{1}{N} \right) \right] = \nonumber \\
= &\sum_n \Big[ \left( A_{n,S} + \frac{\alpha_{n,S}}{N} \right) e^{ - \left( M_{n,S} + \frac{\mu_{n,S}}{N} \right) t } + \nonumber \\
&\left( -A_{n,S} + \frac{\alpha_{n,S}}{N} \right) e^{ - \left( M_{n,S} - \frac{\mu_{n,S}}{N} \right) t } + \nonumber \\
&\left( A_{n,AS} + \frac{\alpha_{n,AS}}{N} \right) e^{ - \left( M_{n,AS} + \frac{\mu_{n,AS}}{N} \right) t } + \nonumber \\
&\left( -A_{n,AS} + \frac{\alpha_{n,AS}}{N} \right) e^{ - \left( M_{n,AS} - \frac{\mu_{n,AS}}{N} \right) t } \Big] = 0
\end{flalign}
This equation must be valid for every value of $N$ and $t$. We want to study the possible solutions. The strategy is to group all the exponentials containing the same mass in Eq.~\eqref{eq:oddpowers}. The prefactor of every different exponential must then vanish. We will assume that:
\begin{enumerate}
\item no accidental degeneracy exists separately in s/asQCD, both at any finite $N$ and at large-$N$;
\item no accidental degeneracy exists in f2QCD at any finite $N$;
\item no accidental degeneracy exists in f2QCD at large-$N$, other that the one predicted by orientifold planar equivalence.
\end{enumerate}
Assumption 1 implies that $m_{n,S} = M_{n,S} + \mu_{n,S}/N$ must be different from $m_{m,S} = M_{m,S} + \mu_{m,S}/N$ for every $m \neq n$ (and analogously for the $AS$ channel). Assumption 2 implies that $m_{n,S} = M_{n,S} + \mu_{n,S}/N$ must be different at finite $N$ from $m_{m,AS} = M_{m,AS} + \mu_{m,AS}/N$ for every $m$.

Moreover, the non-degeneracy assumptions forbid states with no odd powers of $1/N$. In fact if for some state in the S channel $\mu_{n,S}=0$, assuming no accidental degeneracy, the only way to satisfy Eq.~\eqref{eq:oddpowers} is to require that $\alpha_{n,S}=0$. Therefore the term $A_{n,S} e^{M_{n,S} t}$ containing only even powers of $1/N$ appears in the $S$ correlator. But since the even powers of $1/N$ are equal in the $S$/$AS$ correlators thanks to Eq.~\eqref{eq:S/AScorr}, the same term must appear in the $AS$ correlator violating the assumption 2.

A rapid inspection of Eq.~\eqref{eq:oddpowers} shows that the only way to respect the non-degeneracy assumptions and to have non-vanishing amplitudes is to have that $M_{n,S} + \mu_{n,S}/N = M_{n,AS} - \mu_{n,AS}/N$, which also implies $M_{n,S} - \mu_{n,S}/N = M_{n,AS} + \mu_{n,AS}/N$. In fact the even/odd powers of $1/N$ must be separately equal:
\begin{equation} \label{eq:S/ASmasses}
\begin{split}
&M_{n,S}(1/N) = M_{n,AS}(1/N) \equiv M_n(1/N) \ , \\
&\mu_{n,S}(1/N) = - \mu_{n,AS}(1/N) \equiv \mu_n(1/N) \ .
\end{split}
\end{equation}
The sum of the amplitudes of the degenerate exponentials must vanish. Separating even/odd powers of $1/N$:
\begin{equation} \label{eq:S/ASamplitudes}
\begin{split}
&A_{n,S}(1/N) = A_{n,AS}(1/N) \equiv A_n(1/N) \ , \\
&\alpha_{n,S}(1/N) = - \alpha_{n,AS}(1/N) \equiv \alpha_n(1/N) \ .
\end{split}
\end{equation}

Summarizing, the even-power corrections are the same for s/asQCD while the odd-power corrections have opposite sign in the two theories:
\begin{equation} \label{eq:decomposition}
\begin{split}
&a_{n,S} \left( \frac{1}{N} \right) = A_n \left( \frac{1}{N^2} \right) + \frac{1}{N} \alpha_n \left( \frac{1}{N^2} \right) \ ,\\
&a_{n,AS} \left( \frac{1}{N} \right) = A_n \left( \frac{1}{N^2} \right) - \frac{1}{N} \alpha_n \left( \frac{1}{N^2} \right) \ ,\\
&m_{n,S} \left( \frac{1}{N} \right) = M_n \left( \frac{1}{N^2} \right) + \frac{1}{N} \mu_n \left( \frac{1}{N^2} \right) \ ,\\
&m_{n,AS} \left( \frac{1}{N} \right) = M_n \left( \frac{1}{N^2} \right) - \frac{1}{N} \mu_n \left( \frac{1}{N^2} \right) \ . 
\end{split}
\end{equation}

\begin{figure}
\includegraphics*[width=\columnwidth]{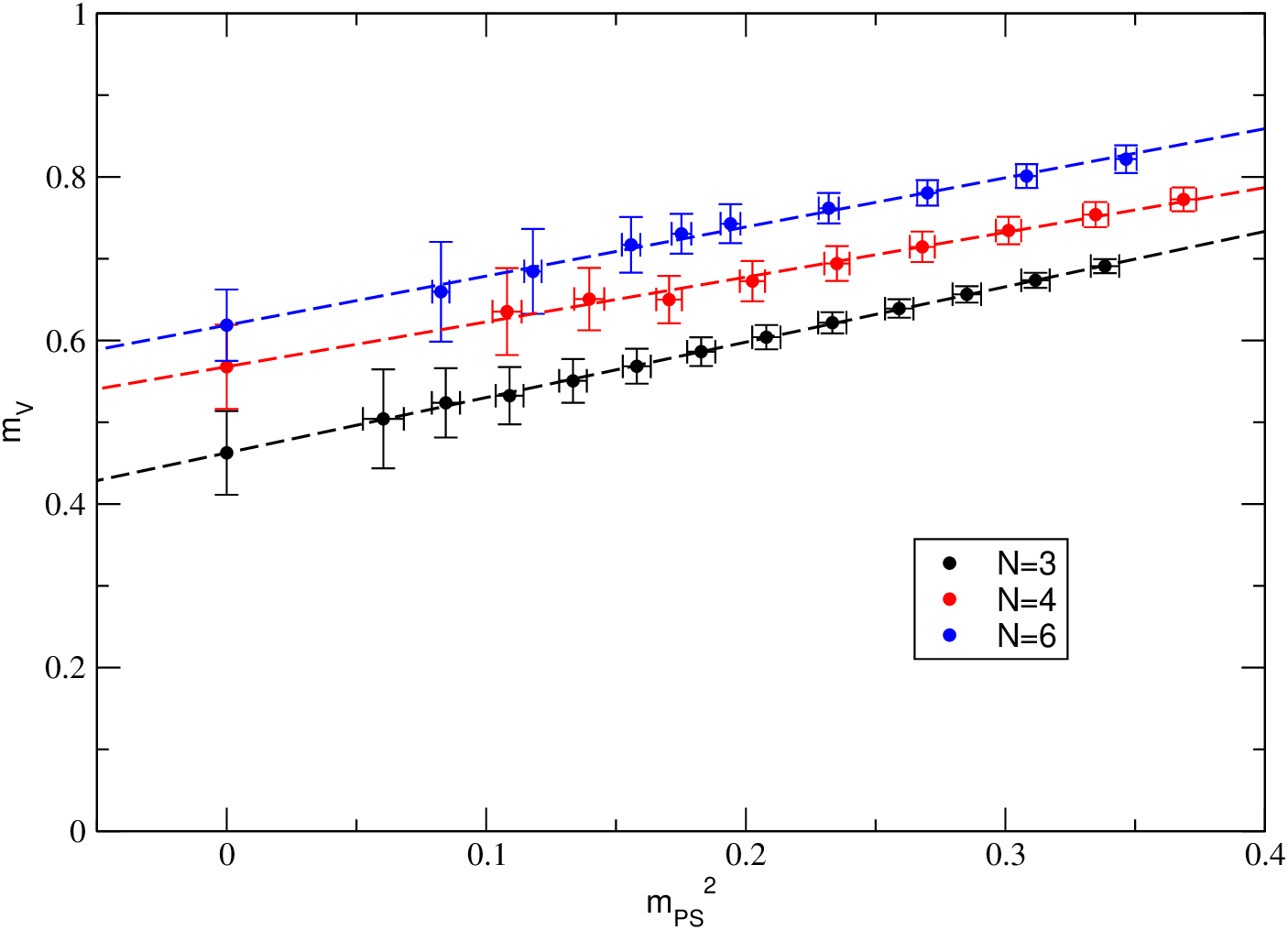}%
\caption{$m_V$ as a function of $m_{PS}^2$ for asQCD. The dashed lines are fits according to Eq.~\eqref{eq:chiral}.\label{fig:asy}}
\end{figure}
\begin{figure}
\includegraphics*[width=\columnwidth]{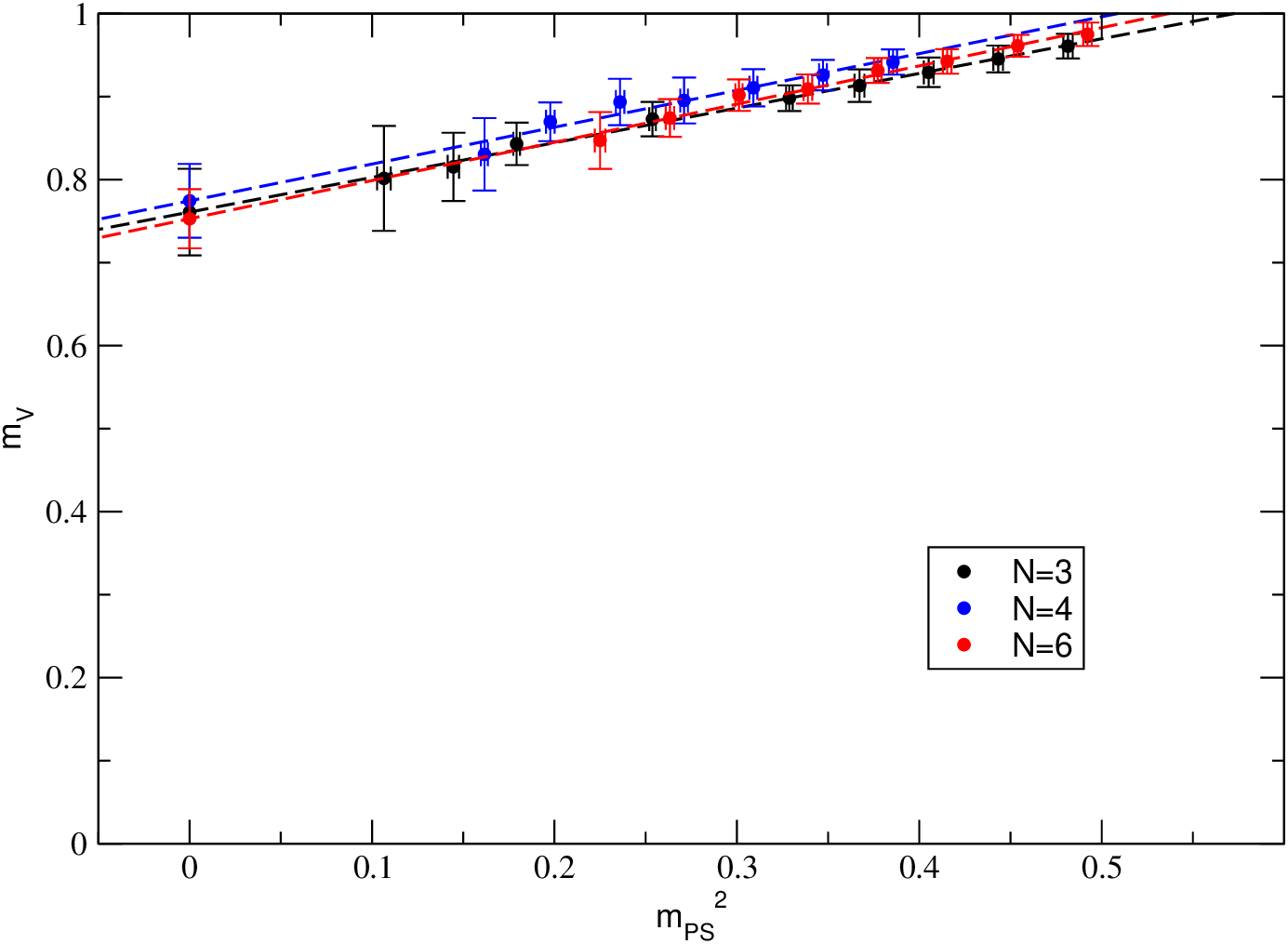}%
\caption{$m_V$ as a function of $m_{PS}^2$ for sQCD. The dashed lines are fits according to Eq.~\eqref{eq:chiral}.\label{fig:sym}}
\end{figure}
\begin{figure}
\includegraphics*[width=\columnwidth]{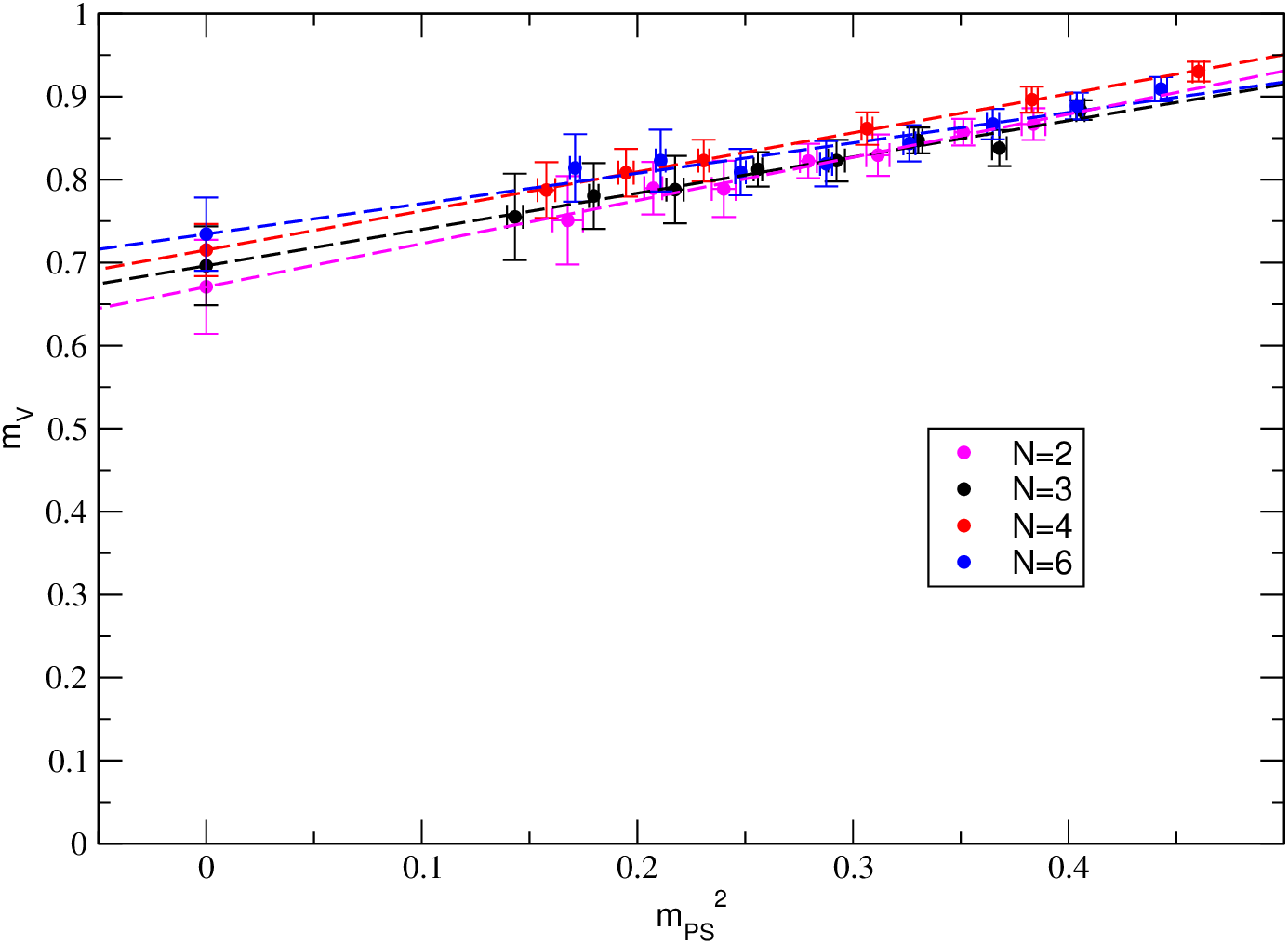}%
\caption{$m_V$ as a function of $m_{PS}^2$ for adjQCD. The dashed lines are fits according to Eq.~\eqref{eq:chiral}.\label{fig:adj}}
\end{figure}

Moving to adjQCD, the normalized correlator is:
\begin{flalign}
\frac{C_{Adj}(t)}{N^2} = & \frac{1}{2} \sum_{\mathbf{x}} \sum_{\omega} c_\omega \frac{\langle |\tr W(\omega)|^2 \rangle_{YM}}{N^2} + \nonumber \\
& - \frac{1}{2N^2} \sum_{\mathbf{x}} \sum_{\omega} c_\omega \ .
\end{flalign}
Since $\langle |\tr W(\omega)|^2 \rangle_{YM}/N^2$ contains only even powers of $1/N$, the same conclusion can be drawn for $C_{Adj}(t)/N^2$. In principle the same mechanism as in f2QCD might happen in the adjQCD, and odd powers of $1/N$ might appear in masses and amplitudes while their absence is preserved in the correlator. This should be due to a pairing of states with masses and amplitude related by the relationships~\eqref{eq:S/ASmasses} and~\eqref{eq:S/ASamplitudes}, and it would imply a degeneracy of states in the large-$N$ limit. While in f2QCD the double degeneracy is explained by orientifold planar equivalence, no special reason exists for such a degeneracy in the large-$N$ limit of adjQCD. We will therefore assume that masses and amplitudes in adjQCD contain only even powers of $1/N$:
\begin{gather}
\frac{C_{Adj}(t)}{N^2} = \sum_n a_{n,Adj} \left( \frac{1}{N^2} \right) \exp \left\{ - t m_{n,Adj} \left( \frac{1}{N^2} \right) \right\} \ .
\end{gather}

In the notation of this section, orientifold planar equivalence translates to the following chains of equalities:
\begin{equation}
\begin{split}
&a_{n,Adj}(0) = a_{n,S}(0) = a_{n,AS}(0) = A_n(0) \ ,\\
&m_{n,Adj}(0) = m_{n,S}(0) = m_{n,AS}(0) = M_n(0) \ .
\end{split}
\end{equation}

\
\begin{table*}[htb]
\setlength{\pbln}{2.1cm}
\begin{tabular}{||c|c|c||c|c|c||c|c|c||}
\hline
\multicolumn{3}{||c||}{SU(3)} & \multicolumn{3}{c||}{SU(4)} & \multicolumn{3}{c||}{SU(6)}\\
\hline
$- am_0$ & \pb{$a m_{PS}$} & \pb{$a m_V$} & $a m_0$ & \pb{$a m_{PS}$} & \pb{$a m_V$} & $a m_0$ & \pb{$a m_{PS}$} & \pb{$a m_V$} \\
\hline
0.81 & 0.5817(47) & 0.6908(86) & 1.09 & 0.6071(41) & 0.772(15) & 1.31 & 0.5886(34) & 0.822(17) \\ 	
0.82 & 0.5582(49) & 0.6736(91) & 1.10 & 0.5786(42) & 0.754(15) & 1.32 & 0.5551(36) & 0.801(15) \\
0.83 & 0.5340(51) & 0.656(10) & 1.11 & 0.5488(44) & 0.734(17) & 1.33 & 0.5196(38) & 0.780(16) \\ 
0.84 & 0.5090(54) & 0.639(11) & 1.12 & 0.5177(47) & 0.714(19) & 1.34 & 0.4816(40) & 0.762(19) \\
0.85 & 0.4830(57) & 0.622(13) & 1.13 & 0.4848(50) & 0.694(21) & 1.35 & 0.4405(43) & 0.743(24) \\ 
0.86 & 0.4559(60) & 0.604(15) & 1.14 & 0.4500(54) & 0.672(25) & 1.355 & 0.4185(45) & 0.730(24) \\
0.87 & 0.4275(63) & 0.586(18) & 1.15 & 0.4127(60) & 0.650(30) & 1.36 & 0.3947(45) & 0.717(34) \\ 
0.88 & 0.3974(68) & 0.568(21) & 1.16 & 0.3736(78) & 0.651(38) & 1.37 & 0.3434(49) & 0.684(52) \\ 
0.89 & 0.3651(73) & 0.550(27) & 1.17 & 0.3285(85) & 0.635(53) & 1.38 & 0.2872(58) & 0.660(61) \\ 
0.90 & 0.3300(81) & 0.532(35) &      & 		  & 	      &      & 		  & 	      \\
0.91 & 0.2904(93) & 0.524(42) &      & 		  & 	      &      & 		  & 	      \\
0.92 & 0.245(16)  & 0.504(60) &      & 		  & 	      &      & 		  & 	      \\
\hline
\end{tabular}

\caption{Numerical results for $m_{PS}$ and $m_V$ in SU(3), SU(4) and SU(6) gauge theories with fermions in the antisymmetric representation.\label{tab:asy}}
\end{table*}

\
\begin{table*}[htb]
\setlength{\pbln}{2.1cm}
\begin{tabular}{||c||c|c||c|c||c|c||}
\hline
     & \multicolumn{2}{c||}{SU(3)} & \multicolumn{2}{c||}{SU(4)} & \multicolumn{2}{c||}{SU(6)} \\
\hline
$-a m_0$ &\pb{$a m_{PS}$}    & \pb{$a m_V$}    & \pb{$a m_{PS}$}    & \pb{$a m_V$}    & \pb{$a m_{PS}$}    & \pb{$a m_V$}     \\
\hline
1.57 & 0.6939(19) & 0.961(15) & 	   & 	       & 	    &          \\
1.58 & 0.6658(20) & 0.945(16) & 	   & 	       & 	    &          \\
1.59 & 0.6365(21) & 0.929(18) & 0.7015(17) & 0.975(14) & 	    &          \\
1.60 & 0.6060(22) & 0.913(20) & 0.6738(15) & 0.961(13) & 	    &          \\
1.61 & 0.5734(15) & 0.898(15) & 0.6444(18) & 0.942(15) & 	    &          \\
1.62 &            &           & 0.6142(16) & 0.931(15) & 0.6209(18) & 0.942(15)\\ 
1.63 & 0.5037(18) & 0.873(21) & 0.5822(20) & 0.909(18) & 0.5893(18) & 0.926(18)\\
1.64 &            &           & 0.5488(18) & 0.902(19) & 0.5559(19) & 0.911(22)\\
1.65 & 0.4233(21) & 0.843(26) & 0.5131(22) & 0.874(23) & 0.5206(20) & 0.895(28)\\
1.66 & 0.3803(41) & 0.815(41) & 0.4743(31) & 0.847(34) & 0.4856(24) & 0.893(28)\\
1.67 & 0.3263(57) & 0.801(63) & 	   &           & 0.4447(26) & 0.870(23)\\
1.68 &  	  & 	      & 	   &           & 0.4020(24) & 0.830(44)\\
\hline
\end{tabular}

\caption{Numerical results for $m_{PS}$ and $m_V$ in SU(3), SU(4) and SU(6) gauge theories with fermions in the symmetric representation.\label{tab:sym}}
\end{table*}

\
\begin{table*}[htb]
\setlength{\pbln}{2.1cm}
\setlength{\pbln}{1.8cm}
\begin{tabular}{||c||c|c||c|c||c|c||c|c||}
\hline
    & \multicolumn{2}{c||}{SU(2)} & \multicolumn{2}{c||}{SU(3)} & \multicolumn{2}{c||}{SU(4)} & \multicolumn{2}{c||}{SU(6)} \\
\hline

$-a m_0$ & \pb{$a m_{PS}$}    & \pb{$a m_V$}    & \pb{$a m_{PS}$}    & \pb{$a m_V$}    & \pb{$a m_{PS}$}    & \pb{$a m_V$}    & \pb{$a m_{PS}$}    & \pb{$a m_V$}    \\
\hline
 1.49 & 0.6194(44) & 0.867(19) &	           &	        & 	     &	         & 	      &	         \\ 
1.50 & 0.5926(33) & 0.857(16) & 0.6368(15) & 0.884(12)  & 0.6784(20) & 0.930(12) & 	      &	         \\
1.51 & 0.5582(48) & 0.829(25) & 0.6065(28) & 0.838(22)  &  	     &	         & 	      &	         \\
1.52 & 0.5285(37) & 0.822(21) & 0.5747(17) & 0.847(16)  & 0.6189(22) & 0.896(16) & 0.6655(21) & 0.909(15)\\	
1.53 & 0.4899(56) & 0.789(34) & 0.5407(36) & 0.823(25)  & 	     &	         & 0.6355(22) & 0.889(16)\\
1.54 & 0.4553(46) & 0.790(31) & 0.5059(20) & 0.812(21)  & 0.5536(23) & 0.861(19) & 0.6040(23) & 0.867(18)\\
1.55 & 0.4094(87) & 0.751(53) & 0.4663(43) & 0.788(41)  & 	     &	         & 0.5711(25) & 0.844(22)\\
1.56 &            &           & 0.4240(26) & 0.780(40)  & 0.4804(26) & 0.823(25) & 0.5363(26) & 0.819(28)\\
1.57 &            &           & 0.3784(49) & 0.755(52)  & 0.4412(40) & 0.808(29) & 0.4979(24) & 0.809(28)\\
1.58 &            &           &            &            & 0.3972(52) & 0.787(34) & 0.4591(26) & 0.823(37)\\
1.58 &            &           &            &            &            &           & 0.4136(28) & 0.814(41)\\
\hline\end{tabular}

\caption{Numerical results for $m_{PS}$ and $m_V$ in SU(2), SU(3), SU(4) and SU(6) gauge theories with fermions in the adjoint representation.\label{tab:adj}}
\end{table*}

\
\begin{table*}[htb]
\setlength{\pbln}{2.1cm}
\setlength{\pbln}{2.1cm}
\begin{tabular}{||c||c|c||c|c||c|c||}
\hline
  & \multicolumn{2}{c||}{$AS$} & \multicolumn{2}{c||}{$S$} & \multicolumn{2}{c||}{$Adj$} \\
\hline
$N$ & \pb{$a m_V^{\chi}$} & \pb{$B$} & \pb{$a m_V^{\chi}$} & \pb{$B$} & \pb{$a m_V^{\chi}$} & \pb{$B$}\\
\hline                     
2 &           &          &           &          & 0.671(57) & 0.52(14)\\
3 & 0.463(51) & 0.68(17) & 0.761(52) & 0.42(14) & 0.696(47) & 0.44(14)\\
4 & 0.568(51) & 0.55(16) & 0.753(36) & 0.460(87) & 0.715(31) & 0.471(88\\
6 & 0.619(43) & 0.60(17) & 0.774(45) & 0.443(15) & 0.734(44) & 0.37(13)\\
\hline
\end{tabular}

\caption{Fit results for the chiral extrapolation of $m_{V}$ according to Eq.~(\ref{eq:chiral}).\label{tab:fits}}
\end{table*}

\section{Results}
\label{sect:numerical}
The measured masses are listed in Tabs.~\ref{tab:asy}-\ref{tab:adj}, with the results of the chiral extrapolation reported in Tab.~\ref{tab:fits}. For every representation and number of colors, the chiral extrapolation has been obtained by fitting (via a bootstrap procedure) the $PS$ and $V$ masses according to Eq.~(\ref{eq:chiral}). The numerical data and the results of the chiral extrapolation are also shown in Figs.~\ref{fig:asy}-\ref{fig:adj}.

We focus on the chiral limit of the $V$ mass as a function of the number of colors. For fermions in the adjoint representation, a fit including only the first subleading $1/N^2$ correction has been performed. The result is
\begin{gather}
a m_{V,Adj}^\chi = 0.736(17)  - \frac{0.28(11)}{N^2} \ .
\end{gather}

For the $S$/$AS$ channels, we use the formulae in~\eqref{eq:decomposition} and we define:
\begin{equation}
\begin{split}
&M = \frac{m_{V,S}^\chi+m_{V,AS}^\chi}{2} \ , \\
&\mu = N \frac{m_{V,S}^\chi-m_{V,AS}^\chi}{2} \ .
\end{split}
\end{equation}

\begin{figure}
\includegraphics*[width=\columnwidth]{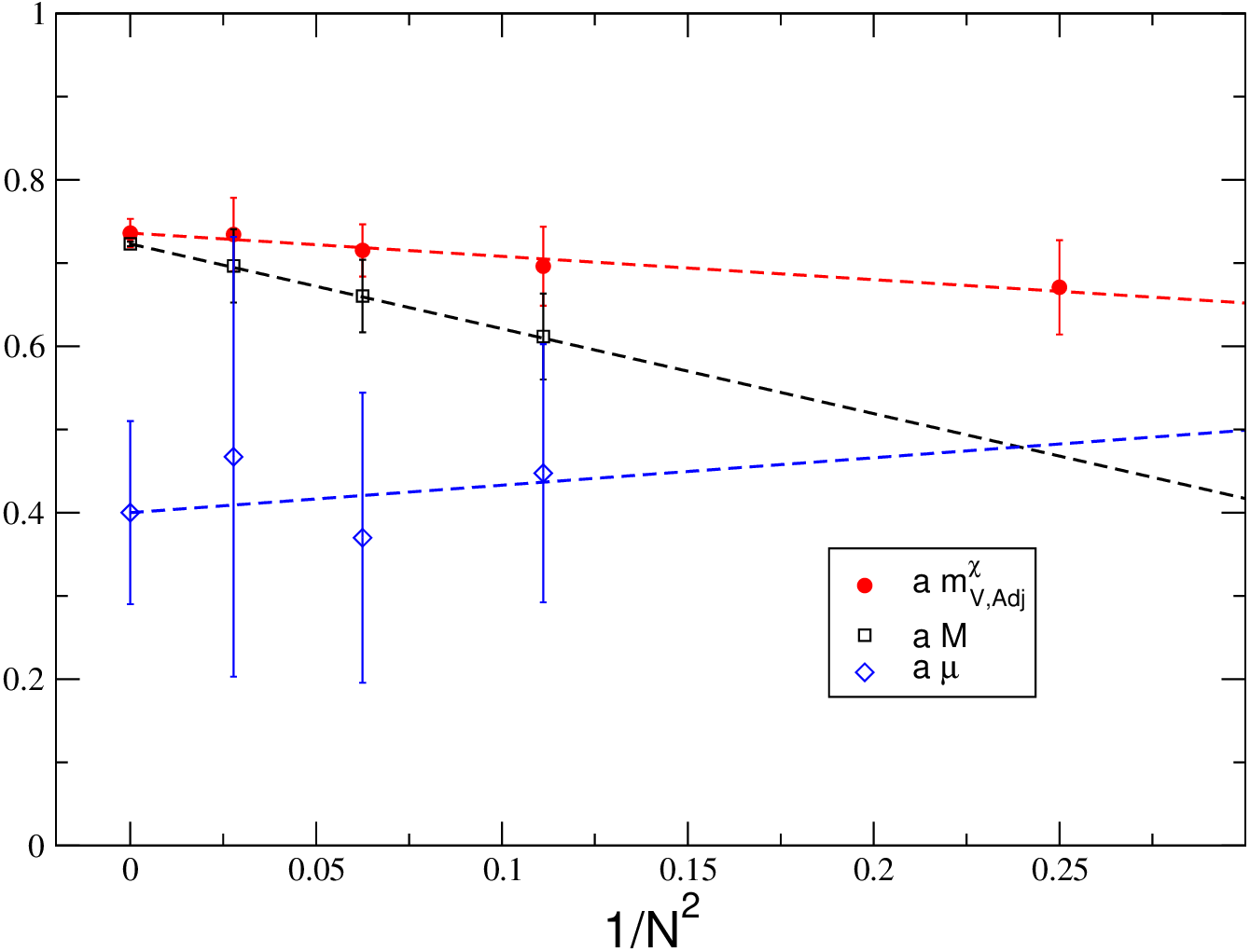}%
\caption{Numerical results for the quantities shown in the legend. The dashed lines are large-$N$ fits assuming corrections of ${\cal O}(1/N^2)$.\label{fig:mmu}}
\end{figure}
\begin{figure}
\includegraphics*[width=\columnwidth]{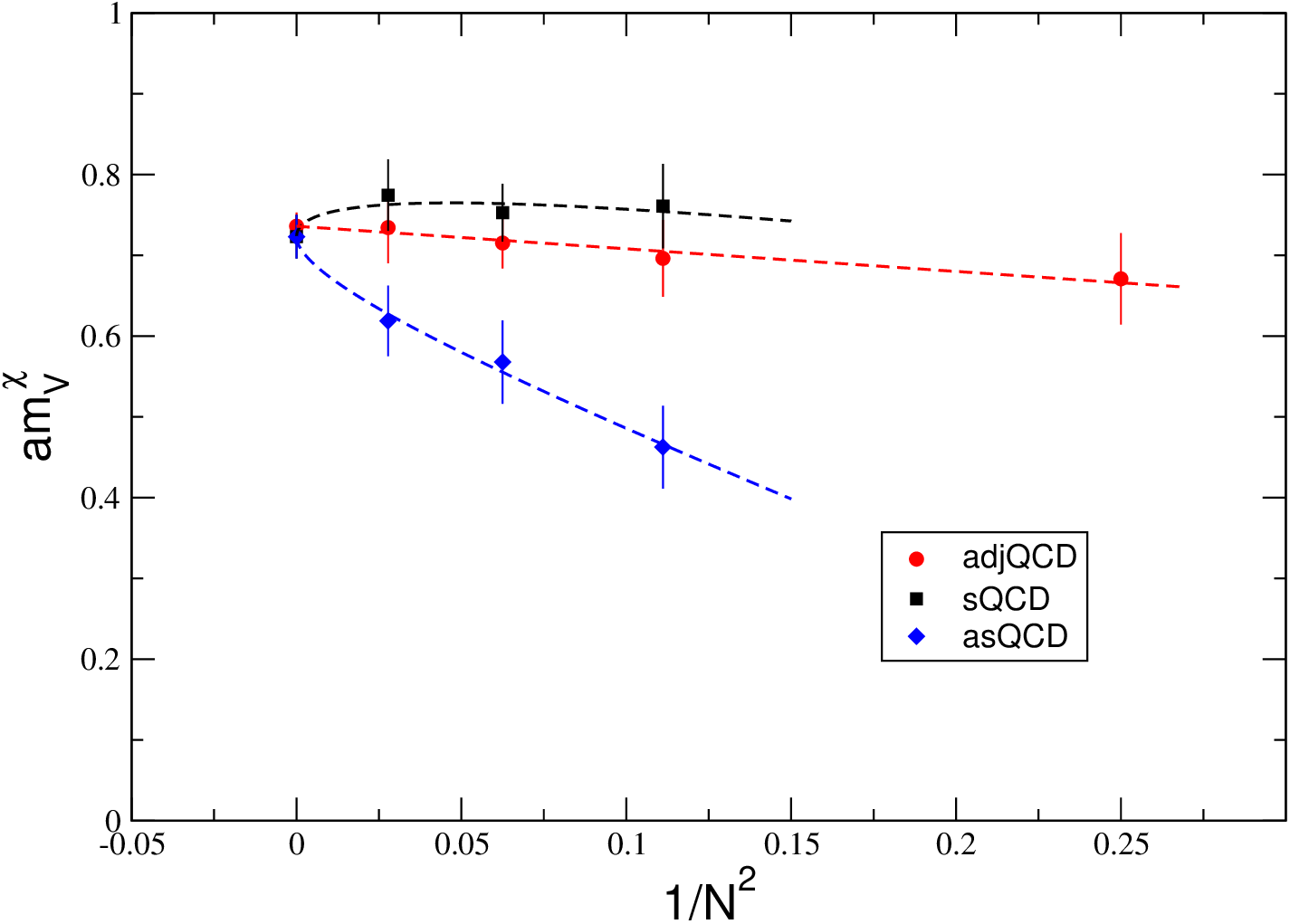}%
\caption{$a m_V^{\chi}$ as a function of $1/N^2$ for fermions in the adjoint, in the symmetric and in the antisymmetric representations. The dashed lines are large-$N$ extrapolations of the numerical data (see text for details).\label{fig:masses_rep}}
\end{figure}

As discussed in Sect.~\ref{sect:corrections}, these two functions contain only even powers of $1/N$. We fit both $M$ and $\mu$ including only the first subleading $1/N^2$ correction, and then we reconstruct the $V$ mass in the $S$/$AS$ channels. We get
\begin{gather}
a m_{V,S}^\chi = 0.723(27)  + \frac{0.40(11)}{N} - \frac{1.02(32)}{N^2} + \frac{0.3(1.2)}{N^3} \ , \\
a m_{V,AS}^\chi = 0.723(27)  - \frac{0.40(11)}{N} - \frac{1.02(32)}{N^2} - \frac{0.3(1.2)}{N^3} \ .
\end{gather}
Note that all coefficients are at most of order one, and the coefficient of the $1/N^3$ term is compatible with zero, which means that our numerical results are not accurate enough to constrain the ${\cal O}(1/N^3)$ term. At low $N$, all terms in the series become equally important, producing a cancellation in the $AS$ channel between the ${\cal O}(1/N^0)$ term and the sub-leading terms. This phenomenon has to be expected, since at $N=2$ the $AS$ representation is the singlet, the $V$ mass vanishes in the chiral limit, and the structure of the terms must be such that they sum to zero. If one wanted to separate the $1/N$ corrections just by looking at the $AS$ channel, large values for $N$ would be needed for which no cancellation happens. The reason the $1/N$ corrections can be obtained by using values of $N$ not larger than 6 deeply relies on the combined analysis we perform of the $S$/$AS$ channels. The same mechanism was observed in the computation of the chiral condensate in~\cite{Armoni:2008nq}.

Our data and fits for $a m_{V,Adj}^\chi$, $a M$ and $a \mu$ are reported in Fig.~\ref{fig:mmu}, while a similar plot for $a m_{V,Adj}^\chi$, $a m_{V,S}^\chi$ and $a m_{V,AS}^{\chi}$ is shown in Fig.~\ref{fig:masses_rep}.

A similar procedure can be used to fit the coefficient B in~\eqref{eq:chiral}, for which we obtain
\begin{gather}
B_{Adj} = 0.405(40) + \frac{0.48(33)}{N^2} \ , \\
B_{S} = 0.495(55)  - \frac{1.3(3.7)}{N} + \frac{0.37(65)}{N^2} - \frac{0.20(32)}{N^3} \ , \\
B_{AS} = 0.495(55)  + \frac{1.3(3.7)}{N} + \frac{0.37(65)}{N^2} + \frac{0.20(32)}{N^3} \ .
\end{gather}All the fits performed for determining coefficients in the large-$N$ expansion have a reduced $\chi^2$ of at most 0.2. Once again, the coefficients of the large-$N$ series come at most of order one, but, differently from the case of the large-$N$ extrapolation of the vector mass in the chiral limit, now only the leading term in the series is different from zero within two standard deviations.

Putting together the large-$N$ extrapolations for $m_{V}^{\chi}$ and $B$, the leading order chiral ansatz for $m_V$ at $N = \infty$ can be parametrized as follows:
\begin{gather}
\label{eq:chiraladj}
a m_{V,Adj} =  0.736(17) + 0.405(40) (a m_{PS,Adj})^2 \ , \\
\label{eq:chiralsas}
a m_{V,S/AS} = 0.723(27) + 0.495(55) (a m_{PS,S/AS})^2 \ .
\end{gather}
The equality of Eqs.~\eqref{eq:chiraladj}~and~\eqref{eq:chiralsas} within errors is a numerical proof of orientifold planar equivalence in the chiral region. We have verified orientifold planar equivalence
with a numerical precision of about 5\%.

\section{Conclusions}
\label{sect:conclusions}
In this work, we have studied orientifold planar equivalence for the pseudoscalar and vector mesons in the quenched theory. By using $N$ up to $6$, we have extrapolated our numerical results for the quenched masses to the large-$N$ limit. While the extrapolation is straightforward for the adjoint representation, the appearance of corrections in odd powers of $1/N$ dictates a combined strategy for the extrapolation of the results in the symmetric and antisymmetric representations. The technique we have discussed allows us to extract the large-$N$ limit chiral behavior of $m_V$ from simulations up to a moderately low $N$. This is an important result in the view of performing dynamical simulations, where the computational cost grows as $N^3$. 

The central quantitative result of our investigation is summarized by Eqs.~\eqref{eq:chiraladj}~and~\eqref{eq:chiralsas}, which shows the equality of the mass of the vector meson in the chiral region at fixed mass of the pseudoscalar meson in the adjoint and in the symmetric/antisymmetric representations. Orientifold planar equivalence was originally introduced as a tool to investigate qualitatively and hopefully quantitatively the nonperturbative behavior of QCD. Assuming that our results are representative of the unquenched case, our data show that the case $N=3$ antisymmetric, which is real-world QCD, is numerically far from its large-$N$ limit, but it is analytically close to it, in the sense that the $N=3$ theory can be obtained from $N=\infty$ by a power series in $1/N$ with coefficients of order one. With the precision of our numerical simulations, which is of the order of a few percent, in order to describe the physics at $N=3$ it is enough to truncate the series at order $1/N^3$. Similar results were obtained for the chiral condensate in~\cite{Armoni:2008nq}. 


\begin{acknowledgements}
We thank Luigi Del Debbio and Claudio Pica for the collaboration that lead to the development of the HiRep Monte Carlo code used in this work. We would also like to thank Adi Armoni for stimulating discussions about orientifold planar equivalence. Numerical simulations have been performed on a 120 core Beowulf cluster partially funded by the Royal Society and STFC and on the IBM Blue/C system at Swansea University. The work of B.L. is supported by the Royal Society through the University Research Fellowship scheme and by STFC under contract ST/G000506/1. A.R. thanks the Deutsche Forschungsgemeinschaft for financial support. A.P. was supported by the European Community - Research Infrastructure Action under the FP7 ``Capacities'' Specific Programme, project ``HadronPhysics2''.
\end{acknowledgements}

\bibliography{orientifold}

\end{document}